\documentclass[12pt]{iopart}

\usepackage{graphicx}
\usepackage{epsf, subfigure, verbatim}
\begin{document}

\title[]{Classical and SUSY solutions of the Boiti-Leon-Manna-Pempinelli equation.}

\author{Laurent DELISLE and Masoud MOSADDEGHI}

\address{D\'epartement de math\'ematiques et de statistique, Universit\'e de Montr\'eal,\\
C.P. 6128, Succc. Centre-ville, Montr\'eal, (QC) H3C 3J7, Canada}
\ead{delisle@dms.umontreal.ca}
\begin{abstract}
In this paper, we propose the study of the Boiti-Leon-Manna-Pempinelli equation from two point of views: the classical and supersymmetric cases. In the classical case, we construct new solutions of this equation from Wronskian formalism and Hirota method. We, then, introduce a $\mathcal{N}=1$ supersymmetric extension of the Boiti-Leon-Manna-Pempinelli equation. We thus produce a bilinear form and give multisolitons and superpartner solutions. As an application, we produce B\"acklund transformations. 

\end{abstract}

\maketitle

\section{Introduction}
Soliton equations has been largely studied in the past in terms of integrability conditions and solutions \cite{Ablo,Kam,Nimmo,Drazin,FOU,Ma,C,Hirota,Mik,Zayed,Luo}. These solutions have been discovered through numerous techniques such as B\"acklund transformations, inverse scattering methods, Lax pairs formulation, $G^{\prime}/G$-expansions, Wronskian formulation and bilinear Hirota method. The later ones will be of interest in this paper. To this instance, the Hirota derivative is defined as
\begin{equation}
\mathcal{D}_x^n(f\cdot g)=(\partial_{x_1}-\partial_{x_2})^nf(x_1)g(x_2)\vert_{x=x_1=x_2}.
\end{equation}

In this paper, we consider the Boiti-Leon-Manna-Pempinelli (BLMP) equation
\begin{equation}
u_{yt}+u_{xxxy}-3u_xu_{xy}-3u_yu_{xx}=0,\label{BLMP}
\end{equation}
where $u=u(x,y,t)$ and subscripts represent partial differentiation with respect to the given variable. This equation has received small attention \cite{Luo} and here we propose to give some new classical solutions through the Hirota and Wronskian methods. To do so, we note that the BLMP equation is strongly related to the Korteweg-de Vries (KdV) equation. Indeed, let us take
\begin{equation}
 u(x,y,t)=p(z,t)=p(x+q(y),t),\label{uform}
\end{equation} 
 where $q$ is an arbitrary function of the variable $y$, then $p$ satisfies
\begin{equation*}
p_{zt}+p_{zzzz}-6p_zp_{zz}=0.
\end{equation*}
Defining $h=p_z$, we see that $h$ satisfies the KdV equation, which has the following bilinear form \cite{Drazin,Hirota,Mik}
\begin{equation}
\mathcal{D}_z(\mathcal{D}_t+\mathcal{D}_z^3)(f\cdot f)=0,\label{biliKdV}
\end{equation}
using the change of variable $h=-2\partial_z^2\log f$. Thus $p=-2\partial_z\log f$ satisfies the BLMP equation for $f(z,t)$ a solution of the above bilinear equation associated to the KdV equation. This suggests that the BLMP equation possesses multisoliton and rational similarity solutions and many other solutions as we will demonstrate.

Without doing the reduction (\ref{uform}) in $u(x,y,t)$  but using the change of variable given above, we can bilinearize, in general, the BLMP equation (\ref{BLMP}) by casting
\begin{equation}
u(x,y,t)=-2\partial_x\log f(x,y,t)-m(y),\label{bilinearfunction}
\end{equation}
 where $m=m(y)$ is an arbitrary function of $y$. One then gets the bilinear equation
\begin{equation}
(\mathcal{D}_y(\mathcal{D}_t+\mathcal{D}_x^3)+3m^{\prime}(y)\mathcal{D}_x^2)(f\cdot f)=0.\label{genbili}
\end{equation}

After considering the classical analysis of the BLMP equation, it is natural to ask what would be its supersymmetric (SUSY) counterpart. SUSY extensions of nonlinear partial differential equation (PDE) has, in the past few decades, been largely studied in terms of integrability and solutions, see for example \cite{Mc,Ibort,Ayari,Carstea,Cars,Ghosh2,Ghosh,Ghosh1,ZLW,Kiselev,FanHon,Delisle,Hussin,Popo,LiuXie,LiuHu,LM}. Here, we consider for the first time a $\mathcal{N}=1$ SUSY extension of the BLMP equation (\ref{BLMP}).

To produce a $\mathcal{N}=1$ SUSY extension of the BLMP equation, we extend, as usual, the space $(x,y,t)$ to the superspace $(x,y,t;\theta)$ \cite{Cornwell} where $\theta$ is a Grassmann variable which satisfies $\theta^2=0$. As usual, we consider a fermionic superfield $\Phi(x,y,t;\theta)$ to get non trivial SUSY extensions \cite{Mc,Ibort,Carstea,Cars,Ghosh}. A $\mathcal{N}=1$ SUSY extension of the BLMP equation (\ref{BLMP}) is given as
\begin{equation}
D_y^2\Phi_t+D_y^2D_x^6\Phi-3D_xD_y(D_x^3\Phi D_yD_x\Phi)=0,\label{SBLMP}
\end{equation}
where $D_x$ and $D_y$ denotes the supercovariant derivatives defined as
\begin{equation}
D_x=\partial_{\theta}+\theta\partial_x,\quad D_y=\partial_{\theta}+\theta\partial_y,\quad D_x^2=\partial_x,\quad D_y^2=\partial_y.\label{Cov}
\end{equation}
Using the finite Taylor expansion
\begin{equation}
\Phi(x,y,t;\theta)=\xi(x,y,t)+\theta u(x,y,t),
\end{equation}
where $\xi$ is a complex-valued fermionic (odd) function and $u$ is a complex-valued bosonic (even) function, we get a set of two equations
\begin{eqnarray}
u_{yt}+u_{xxxy}-3u_{xx}u_y-3u_xu_{xy}+3\xi_x\xi_{xxx}=0,\\
\xi_{yt}+\xi_{xxxy}-3u_{xy}\xi_x-3u_x\xi_{xy}=0.
\end{eqnarray}
In the fermionic limit where $\xi$ is set to zero, we retrieve the classical BLMP equation as expected.

As in the classical case, the SUSY BLMP equation is related to the SUSY KdV equation. Indeed, supposing that $\xi(x,y,t)=\sqrt{q^{\prime}(y)}\chi(z,t)$ and $u(x,y,t)=v(z,t)$ where $z=x+q(y)$, then the above set of equations leads to
\begin{eqnarray}
v_{zt}+v_{zzzz}-6v_zv_{zz}+3\chi_z\chi_{zzz}=0,\label{eqn11}\\
\chi_t+\chi_{zzz}-3\chi_zv_z=0.\label{eqn22}
\end{eqnarray}
In the fermionic limit, we get the classical KdV equation for the variable $v_z$. The equations (\ref{eqn11}) and (\ref{eqn22}) becomes
the SUSY KdV equation first proposed by Manin and Radul \cite{Manin} after identifying $\chi_z=\varphi$. We thus get a series of solutions of the SUSY BLMP equation using this procedure.

In section 2, we give new classical solutions of the BLMP equation using the Wronskian and Hirota bilinear formulation. In section 3, we introduce the SUSY Bell polynomials which will be the key ingredient of our analysis. In section 4, we give solutions of the SUSY BLMP equation, \textit{e.g.} soliton and periodic solutions. There is also an attempt to find the superpartner solutions of the given classical solutions $u$ of the BLMP equation. In section 5, we produce  bilinear B\"acklund transformations using the SUSY Bell polynomials described in section 3.

\section{New classical solutions}

In the introduction, we have shown a link between the BLMP equation and the KdV equation. It means that, for example, we can produce rational similarity solutions for the BLMP equation based on the corresponding ones for KdV. Indeed, for the rational similarity solutions \cite{Ablo,Kam,Drazin,FOU,C,Zayed}, we get
\begin{equation*}
u_n(x,y,t)=\phi_n(z,t)=-2\frac{d}{dz}\log\mathcal{Q}_n(z(3t)^{-\frac{1}{3}}),
\end{equation*}
 where the functions $\mathcal{Q}_n$ are known as the Yablonskii-Vorob'ev polynomials and are defined recursively \cite{C} as
 \begin{equation*}
 \mathcal{Q}_{n+1}\mathcal{Q}_{n-1}=z\mathcal{Q}_n^2-4(\mathcal{Q}_n
 \mathcal{Q}_{n,zz}-\mathcal{Q}_{n,z}^2),
\end{equation*}
with $\mathcal{Q}_0(z)=1$ and $\mathcal{Q}_1(z)=z$. For example, we get the solutions 
\begin{equation*}
u_2(z,t)=-\frac{6z^2}{12t+z^3},\quad u_3(z,t)=\frac{12(30tz^2+z^5)}{720t^2-60tz^3-z^6}.
\end{equation*}
This infinite series of rational similarity solutions have not been obtained before.

We can exhibit soliton solutions. Indeed, for the $N$-soliton solution one gets for the function $f$ \cite{Ablo,Hirota}
\begin{equation}
f_N=\sum_{\mu=0,1}\exp\left(\sum_{i=1}^N\mu_i\psi_i+\sum_{i<j}
a_{ij}\mu_i\mu_j\right),\label{fN}
\end{equation}
where $\psi_i=\kappa_i z-\kappa_i^3t$ and $e^{a_{ij}}=A_{ij}=\left(\frac{\kappa_i-\kappa_j}{\kappa_i+\kappa_j}\right)^2$. The above solutions has been deduced from the reduction of the BLMP equation to the KdV equation via (\ref{uform}). 

Using a Wronskian formulation for the KdV equation \cite{Nimmo,Ma,Hirota}, we get rational, positon, negaton and complexiton solutions of the BLMP equation. To this instance, we look for solutions $f=f(z,t)$ of the following form
\begin{equation}
f=\det(H^{(0)}, H^{(1)},\cdots , H^{(N-1)}),
\end{equation}
where $H=H^{(0)}=(h_1,h_2,\cdots, h_N)^T$ and $H^{(i)}=\frac{\partial^i}{\partial z^i}H$ for $i=1,2,\cdots, N-1$. The resulting function $u=-2\partial_z\log f$ is called a Wronskian solution of order $N$. The Wronskian solution requires \cite{Nimmo,Ma, Hirota,Luo} that the functions $h_i$ satisfy the following linear PDE's
\begin{equation}
-h_{i,zz}=\lambda_i h_i,\quad h_{i,t}=-4 h_{i,zzz},
\end{equation}
for $i=1,2,\cdots N$. When $\lambda_i$ is real, we get three cases of solutions: rational ($\lambda_i=0$), positon ($\lambda_i>0$) and negaton ($\lambda_i<0$). For the complexiton solution, the eigenvalues $\lambda_i$ are supposed to be complex.

 Here we give some examples and refer the interested readers to \cite{Nimmo,Ma,Hirota,Luo} for further discussions on the Wronskian solutions.

A negaton solution of second order
\begin{equation}
u_n(z,t,\gamma_1)=\frac{8\gamma_1\cosh^2(\gamma_1(z-4\gamma_1^2t))}{2\gamma_1(12\gamma_1^2t-z)-\sinh(2\gamma_1(z-4\gamma_1^2t))},\label{un}
\end{equation}
a positon of second order
\begin{equation}
u_p(z,t,\gamma_1)=-\frac{8\gamma_1\cos^2(\gamma_1(z+4\gamma_1^2t))}{2\gamma_1(12\gamma_1^2t+z)+\sin(2\gamma_1(z+4\gamma_1^2t))}\label{up}
\end{equation}
and a complexiton
\begin{equation}
u_c(z,t,\eta)=\frac{4i\hbox{Im}(\eta)\cosh(\bar{\chi})\cosh(\chi)}{\sqrt{\eta}\cosh(\bar{\chi})\sinh(\chi)-\sqrt{\bar{\eta}}\cosh(\chi)
\sinh(\bar{\chi})}\label{uc},
\end{equation}
\begin{figure}[h!]
\centering
\includegraphics[width=2in]{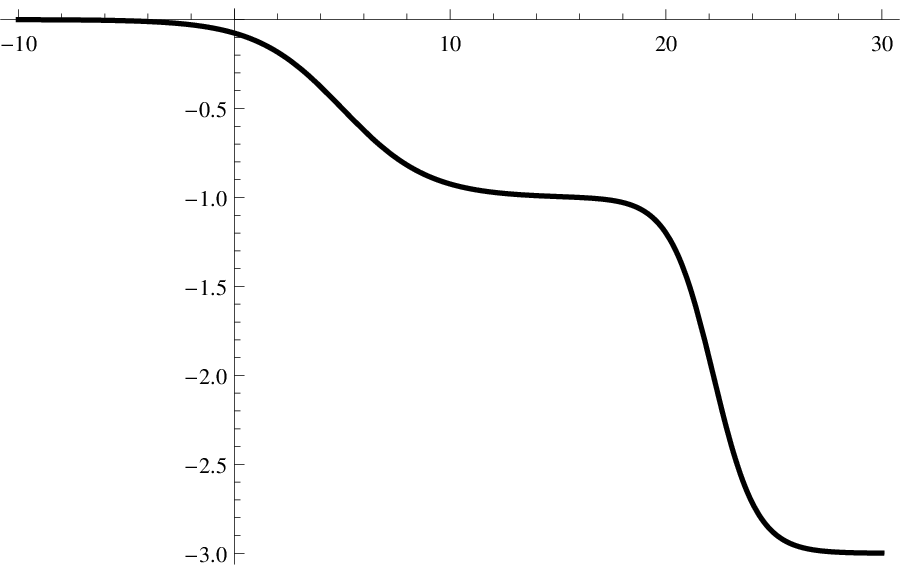}
\includegraphics[width=2in]{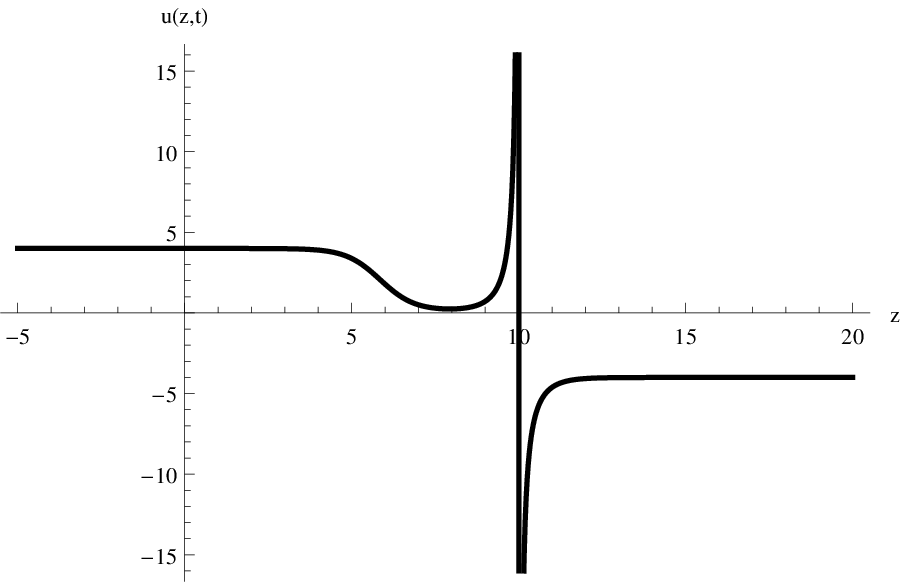}
\includegraphics[width=2in]{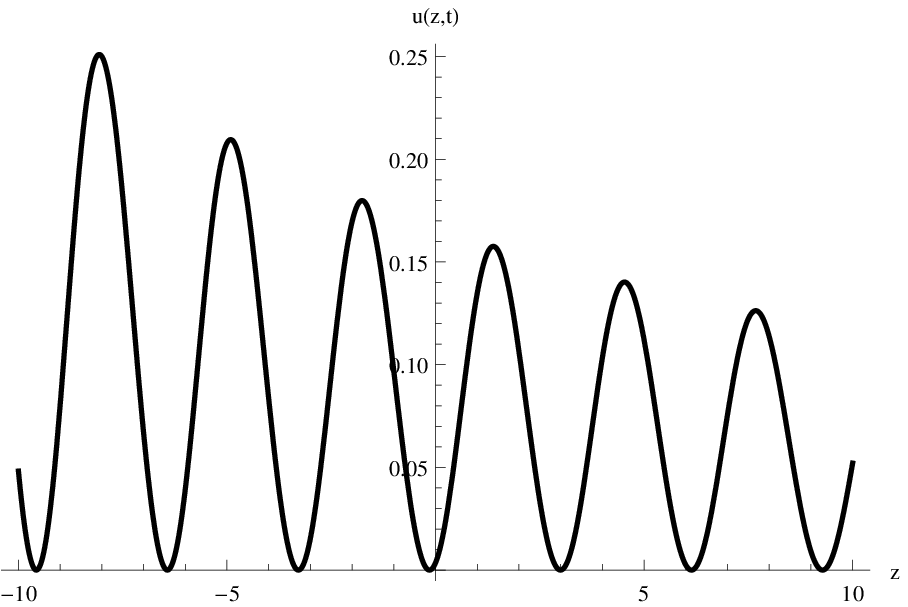}
\caption{From left to right, a 2-soliton solution (\ref{fN}) for $N=2$ ($\kappa_1=\frac12$, $\kappa_2=1$) and $t=20$, a negaton of second order (\ref{un}) ($\gamma_1=1$) and a positon of second order (\ref{up}) ($\gamma_1=1$) for $z=x+q(y)$ and $t=2$.}
\end{figure}
where $\chi=\sqrt{\eta}(z-4\eta t)$, $\bar{\chi}=\sqrt{\bar{\eta}}(z-4\bar{\eta} t)$ and $\eta=\alpha+i\beta$. There is numerously more solutions to the BLMP equation such as interacting solutions. For example, one can construct a rational-soliton ($u_{rs}$) and rational-positon ($u_{rp}$) solutions using Wronskian formulation \cite{Nimmo,Ma, Hirota,Luo}. One gets
\begin{equation}
u_{rs}(z,t,\gamma_1)=\frac{-2\gamma_1^2z}{\gamma_1z\tanh(\gamma_1(z-4\gamma_1^2t))-1}\label{urs}
\end{equation}
and
\begin{equation}
u_{rp}(z,t,\gamma_1)=\frac{-2\gamma_1^2z}{\gamma_1z\tan(\gamma_1(z+4\gamma_1^2t))+1}\label{urp}
\end{equation}
\begin{figure}[h!]
\centering
\includegraphics[width=2in]{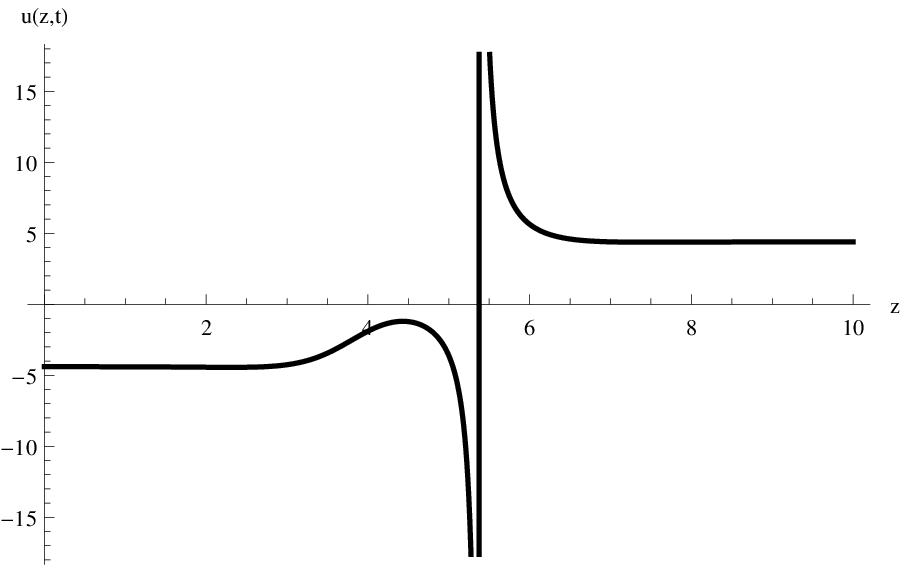}
\includegraphics[width=2in]{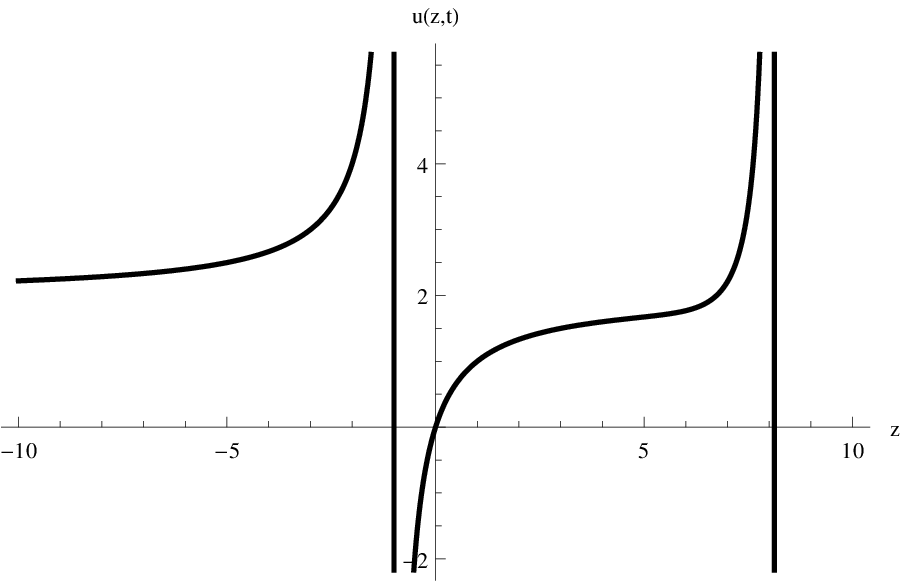}
\includegraphics[width=2in]{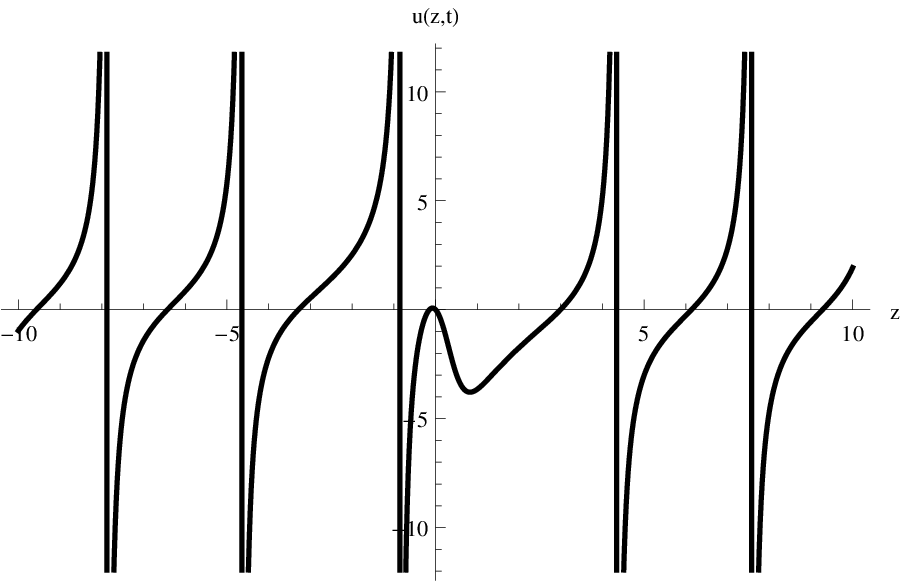}
\caption{From left to right, a complexiton (\ref{uc}) ($\alpha=\beta=1$), a rational-soliton (\ref{urs}) and a rational-positon (\ref{urp}) solutions for $z=x+q(y)$, $t=2$ and $\gamma_1=1$.}
\end{figure}
In the general case, one could be interested in solving the bilinear equation (\ref{genbili}) and retrieve, for example, a general travelling wave solution. Indeed, if the following constraints are supposed
\begin{equation*}
f=f(w)=f\left(a x+\frac{1}{a}m(y)-4a^3t\right),\quad f_{ww}=\alpha f_{w}+\frac{1}{4}(1-\alpha^2)f,
\end{equation*}
where $\alpha$ is an arbitrary complex constant, then the bilinear form (\ref{genbili}) is automatically satisfied and we get a solution of the BLMP equation
\begin{equation}
u=-2a\partial_{w}\log\left(c_1 e^{\frac{1}{2}(1+\alpha)w}+c_2 e^{\frac{1}{2}(-1+\alpha)w}\right)-m(y),\label{trav}
\end{equation}
where $c_1$ and $c_2$ are integration constants and $m=m(y)$ is an arbitrary function of the variable $y$ used in (\ref{bilinearfunction}) to bilinearize the BLMP equation . At our knowledge, the above solution is new. Here we give an example for $\alpha=0$ and $c_1=c_2=1$, one gets the kink-type solution
\begin{equation*}
u(x,y,t)=-a \tanh(\frac{w}{2})-m(y).
\end{equation*}

Most of the above solutions are new and will be useful in subsequent sections to construct superpartners of the supersymmetric BLMP equation.

\section{SUSY extension of the Bell polynomials}
The Bell polynomials are known as an alternative to tedious calculations of finding a bilinear form of an equation \cite{Luo,FanHon}. Indeed, we are able to rewrite the equation in terms of the Bell polynomials and to get simultaneously the change of variable and the bilinear form of the studied equation. For more details, we refer the reader to \cite{FanHon}.

For our present analysis, we introduce a particular case of the SUSY Bell polynomials. We refer the interested reader to \cite{FanHon} for a more general introduction to the SUSY Bell polynomials. 

Let $A(X;\theta)$ be a superdifferentiable bosonic superfield \cite{Cornwell} where $X=(x,y,t)$. The super Bell polynomials are defined as
\begin{equation}
Y_{l\cdot X,(k_x,k_y)}(A)=Y_{l\cdot X,(k_x,k_y)}(A_{l\cdot X,(k_x,k_y)})=e^{-A}D_x^{k_x}D_y^{k_y}\partial_x^{l_x}\partial_y^{l_y}\partial_t^{l_t}e^{A},
\end{equation}
where $l\cdot X=(l_x x, l_y y, l_t t)$ and $D_x$ and $D_y$ are the supercovariant derivatives defined in (\ref{Cov}).

The super Bell polynomials enjoys a remarkable property given by
\begin{equation}
Y_{l\cdot X,(k_x,k_y)}(A)\vert_{A=\ln f}=\frac{f_{l\cdot X,(k_x,k_y)}}{f}:=\frac{D_x^{k_x}D_y^{k_y}\partial_x^{l_x}\partial_y^{l_y}\partial_t^{l_t}f(X;\theta)}{f(X;\theta)}.\label{prop}
\end{equation}

The super binary polynomials are defined, using the super Bell polynomials, as
\begin{equation}
\mathcal{Y}_{l\cdot X,(k_x,k_y)}(v,w)=Y_{l\cdot X,(k_x,k_y)}(A_{s\cdot X,(n_x,n_y)}),
\end{equation}
where the different derivatives of $A$ are replaced by the superfields $v$ and $w$ using the following procedure
\[A_{s\cdot X,(n_x,n_y)} =
\left\{
	\begin{array}{ll}
		v_{s\cdot X,(n_x,n_y)}, \,  \, s_x+s_y+s_t+n_x+n_y\,\, \mbox{is odd} \\
		w_{s\cdot X,(n_x,n_y)}, \,  s_x+s_y+s_t+n_x+n_y\,\, \mbox{is even}
	\end{array}
\right.
\]
As an example, we have that
\begin{eqnarray*}
Y_{(3x,0,0),(0,0)}(A)=A_{xxx}+3A_xA_{xx}+A_x^3,\\ \mathcal{Y}_{(3x,0,0),(0,0)}(v,w)=v_{xxx}+3v_xw_{xx}+v_x^3.
\end{eqnarray*}
The super binary Bell polynomials are related to the SUSY Hirota derivative \cite{FanHon}. Indeed, we have the following relation
\begin{equation}
\mathcal{Y}_{l\cdot X,(k_x,k_y)}\left(v=\ln\frac{f}{g},w=\ln fg\right)=(fg)^{-1}\mathcal{S}_x^{k_x}\mathcal{S}_y^{k_y}\mathcal{D}_x^{l_x}\mathcal{D}_y^{l_y}\mathcal{D}_t^{l_t}(f\cdot g),\label{binBell}
\end{equation}
where the super Hirota derivative is defined as
\begin{equation}
\mathcal{S}_{\mu}(f\cdot g)=(D_{\mu}-D_{\mu^{\prime}})f(X,\theta)g(X^{\prime},\theta^{\prime})\vert_{X=X^{\prime},\theta^{\prime}=\theta}.
\end{equation}

The super $P$-polynomials \cite{FanHon} are a particular class of (\ref{binBell}) where we take $f=g$. We have that
\begin{equation}
P_{l\cdot X,(k_x,k_y)}(w)=\mathcal{Y}_{l\cdot X,(k_x,k_y)}(0,w=2\ln g)=g^{-2}\mathcal{S}_x^{k_x}\mathcal{S}_y^{k_y}\mathcal{D}_x^{l_x}
\mathcal{D}_y^{l_y}\mathcal{D}_t^{l_t}(g\cdot g).
\end{equation}
A direct consequence of (\ref{binBell}) is that $P_{l\cdot X,(k_x,k_y)}(w)=0$ whenever $l_x+l_y+l_t+k_x+k_y$ is odd.

\section{Bilinear form, multisolitons and superpartners}
 In this section,
we propose a bilinear form of the SUSY BLMP equation (\ref{SBLMP}) and give multisolitons and superpartner solutions. In order to get this form, let us cast the change of variable $\Phi=D_{x}F+\Lambda(y;\theta)$ where $F$ is a bosonic superfield, $\Lambda$ is a fermionic superfield and integrate with
respect to $D_{x}$ and secondly by $D_{y}$. We thus get the fermionic equation
\begin{equation}
 D_{y}(F_t+D_{x}^6F)-3(D_{x}^4F) (D_{y}
 D_{x}^2F)-3(
D_{y}D_{x}\Lambda)F_{xx}=0,\label{SSBLMP}
\end{equation}
where the constants of integration as been set to zero and we recall that $D_x^2=\partial_x$. Using the following P-polynomials
\begin{equation*}
 P_{(0,0,t),(0,1)}(w)=D_{y}w_t,\quad P_{(3x,0,0),(0,1)}(w)=D_{y}w_{xxx}+3w_{xx}D_{y}w_x,
\end{equation*}
equation (\ref{SSBLMP}) can be rewritten as the polynomial equation
\begin{equation*}
 P_{(0,0,t),(0,1)}(-F)+P_{(3x,0,0),(0,1)}(-F)-3(
D_{y}D_{x}\Lambda)P_{(2x,0,0),(0,0)}(-F)=0.
\end{equation*}
We thus cast $F=-2\ln g$ and we get the bilinear form of the SUSY BLMP equation (\ref{SBLMP}) given by
\begin{equation}
(\mathcal{S}_{y}\mathcal{D}_t+\mathcal{S}_{y}\mathcal{D}_x^3-3(
D_{y}D_{x}\Lambda)\mathcal{D}_x^2)(g\cdot g)=0.\label{sbilimod}
\end{equation}
Note that if $\Lambda(y;\theta)=\eta(y)+\theta c(y)$ then we get $D_{y}D_{x}\Lambda=\theta c^{\prime}(y)$ and, in terms of components, if $g=g^b+\theta g^f$, the bilinear equation (\ref{sbilimod}) becomes
\begin{eqnarray*}
\mathcal{Q}_1(\mathcal{D}_x,\mathcal{D}_y,\mathcal{D}_t)(g^b\cdot g^b)=(\mathcal{D}_y(\mathcal{D}_t+\mathcal{D}_x^3)-3 c^{\prime}(y)\mathcal{D}_x^2)(g^b\cdot g^b)=0,\\ 
\mathcal{Q}_2(\mathcal{D}_x,\mathcal{D}_y,\mathcal{D}_t)(g^b\cdot g^f)=(\mathcal{D}_t+\mathcal{D}_x^3)(g^b\cdot g^f)=0.
\end{eqnarray*}
In order to find the dispersion relations of $g^b$ and $g^f$, one as to solve the set of equations
\begin{equation*}
 \mathcal{Q}_1(\kappa,\rho,\omega)=0,\quad \mathcal{Q}_2(\kappa,\rho,\omega)=0,
\end{equation*}
which is equivalent to
\begin{equation*}
 \rho(\omega+\kappa^3)-3 c^{\prime}(y)\kappa^2=0,\quad \omega+\kappa^3=0.
\end{equation*}
Thus we get $c^{\prime}(y)=0$. This shows that to get non-trivial multisoliton solutions, one as to consider the super bilinear form
\begin{equation}
 \mathcal{S}_{y}(\mathcal{D}_t+\mathcal{D}_x^3)(g\cdot g)=0.\label{sbili}
\end{equation}

\subsection{Multisolitons}
As usual, in order to get super multisolitons of the SUSY BLMP equation (\ref{SBLMP}), we use the $\epsilon$-expansion of $g$
\begin{equation*}
 g=1+\epsilon g_1+\epsilon^2 g_2+\cdots =\sum_{k=0}^{\infty}\epsilon^kg_k
\end{equation*}
and substitute this expansion in the super bilinear form (\ref{sbili}). For the super travelling wave or one super soliton solution, we take
\begin{equation*}
 g_1=e^{\phi_1},\quad \phi_1=\kappa_1x+\rho_1y+\omega_1t+\theta\zeta_1=\varphi_1+\theta\zeta_1
\end{equation*}
and $g_i=0$ for all $i\geq2$. Introducing $g$ in the super bilinear form (\ref{sbili}), we get the dispersive relation $\omega_1=-\kappa_1^3$. For the two super soliton solution, we take
\begin{equation}
g_1=e^{\phi_1}+e^{\phi_2},\quad g_2=A_{12}\left(1+2\frac{\zeta_1\zeta_2}{\rho_2-\rho_1}\right)e^{\varphi_1+\varphi_2+\theta(\zeta_1 \alpha_{12}+\zeta_2 \alpha_{21})}
\end{equation}
and $g_i=0$ for all $i\geq 3$. In this case, we get the even relations for $i=1,2$
\begin{equation}
\omega_i=-\kappa_i^3,\quad A_{12}=\frac{(\kappa_1-\kappa_2)(\rho_1-\rho_2)}{(\kappa_1+\kappa_2)(\rho_1+\rho_2)},\quad \alpha_{ij}=\frac{\rho_i+\rho_j}{\rho_i-\rho_j}.
\end{equation}

The generalization to super multisoliton solutions is direct following the procedure described in \cite{Cars}. An interesting feature of the above solution is the absence of fermionic constraints.

 We refer the readers to \cite{Mc,Ibort,Carstea,Cars,Ghosh2,Ghosh,Ghosh1,Kiselev,Delisle,Hussin} for more details on the construction of supersymmetric multisoliton solutions.
 
\subsection{Superpartners}
In section 2, we have presented numerous classical solutions of the BLMP equation (\ref{BLMP}). Here we get for some of them, the superpartner $\xi=\zeta k$, where $\zeta$ is an odd constant and $k=k(x,y,t)$ using $\Phi=\xi+\theta u$.
 
 Let us consider solution (\ref{trav}) which takes the form 
\begin{equation}
 u_{(d_1,d_2)}=-a(1+\alpha)+\frac{2a(d_1-d_2)}{d_1-d_2+(d_1+d_2)e^w}-m(y),
 \end{equation}
where $d_1=c_1+c_2$ and $d_2=c_1-c_2$. In this case, defining $\psi=\xi_{w}$, we see that $\psi=\psi(w)$ satisfies the time independent Schr\"odinger equation \cite{Drazin,Hirota,Mik}
 \begin{equation}
 \psi_{ww}=\left(4-\frac{3(d_1^2-d_2^2)}{2(d_1\cosh(\frac{w}{2})+d_2\sinh(\frac{w}{2}))^2}\right)\psi.\label{Schro}
 \end{equation}
It can be solved using the change of variable $\mu=\tanh(\frac{w}{2})$. We get the solution $\xi=\zeta k$ where $\zeta^2=0$ and
 \begin{eqnarray*}
 k_{(d_1,d_2)}(w)=\beta_1(4d_1d_2(11d_1^2+d_2^2)\cosh\frac{w}{2}+15d_1d_2(d_1^2-d_2^2)\cosh\frac{3w}{2}\\+5d_1d_2(d_2^2-d_1^2)\cosh\frac{5w}{2}+4d_2^2(11d_1^2+d_2^2)\sinh\frac{w}{2}\\
 -5(d_1^2-d_2^2)(4d_1^2-d_2^2)\sinh\frac{3w}{2}-(d_1^2-d_2^2)(4d_1^2+d_2^2)\sinh\frac{5w}{2})/\\(d_1\cosh\frac{w}{2}+d_2\sinh\frac{w}{2})+\beta_2(4d_1(5d_1^2-d_2^2)(d_1^2+d_2^2)\cosh\frac{w}{2}+\\
 5d_1(d_1^2-d_2^2)(5d_1^2-3d_2^2)\cosh\frac{3w}{2}+5d_1(d_1^4-d_2^4)\cosh\frac{5w}{2}+\\
 4d_2(5d_1^2-d_2^2)(d_1^2+d_2^2)\sinh\frac{w}{2}+5d_2(d_2^4-d_1^4)\sinh\frac{3w}{2}+\\
 d_2(d_1^2-d_2^2)(9d_1^2+d_2^2)\sinh\frac{5w}{2})/(d_1\cosh\frac{w}{2}+d_2\sinh\frac{w}{2})+\beta_3,
 \end{eqnarray*}
 where the $\beta_i$'s are constants of integration. 
 
Let us give some relevant special solutions. The case $(d_1,d_2)=(1,0)$ (corresponding to the P\"oschl-Teller potential \cite{Pos} in the Schr\"odinger equation), we get the kink-type (soliton) solution
\begin{equation*}
u_{(1,0)}=-a \tanh(\frac{w}{2})-a\alpha-m(y)
\end{equation*}
and the corresponding superpartner $\xi_{(1,0)}=\zeta k_{(1,0)}$ where
\begin{eqnarray*}
k_{(1,0)}(w)=\beta_3-8\beta_1(4\sinh(w)+\sinh(2w)-2\tanh(\frac{w}{2}))\\  +10\beta_2(4\cosh(w)+\cosh(2w)).
\end{eqnarray*}
The case $(d_1,d_2)=(0,1)$ leads to
\begin{equation*}
u_{(0,1)}=-a \coth(\frac{w}{2})-a\alpha-m(y)
\end{equation*}
with corresponding superpartner $\xi_{(0,1)}=\zeta k_{(0,1)}$ where
\begin{equation*}
k_{(0,1)}(w)=\beta_3+2(\beta_1-\beta_2)(-4\cosh(w)+\cosh(2w)).
\end{equation*}
 The above solutions have never been considered and we want to insist on the novelty which is due to a new approach of solving SUSY models.
 
 \section{Bilinear B\"acklund transformations}

B\"acklund transformations \cite{Nimmo,Drazin,Hirota,Mik,Luo,ZLW,Popo,LiuXie,LiuHu} has been shown to be useful in constructing new solutions from known ones and give nonlinear superposition formulas. 

The above sections have exhibited new solutions of the classical and SUSY BLMP equation. Here we propose a nice application of the SUSY Bell polynomials applied to the SUSY case. Indeed, we propose to find bilinear B\"acklund transformations.
To achieve this, we define
\begin{equation*}
L(\Phi)=D_{x}\Phi_{yt}+D_{x}\Phi_{xxxy}-3D_{x}D_{y}(\Phi_{xx}D_{y}\Phi_x+\Phi_{xx}D_{y}D_{x}\Lambda)
\end{equation*}
and we suppose that $\psi=-2\ln\tau$ and $\eta=-2\ln\mu$ are such that $L(\psi)=L(\eta)=0$. In order to use the super binary Bell polynomials, we define new variables
\begin{equation}
 v=-\frac{\psi-\eta}{2}=\ln\left(\frac{\tau}{\mu}\right),\quad w=-\frac{\psi+\eta}{2}=\ln(\tau\mu)\label{change}
\end{equation}
and, in terms of $v$ and $w$, we have the following equation 
\begin{eqnarray*}
L(\psi)-L(\eta)=-2(D_{x}v_{yt}+D_{x}v_{xxxy}\\+3D_{x}D_{y}(v_{xx}D_{y}w_x+w_{xx}D_{y}v_x-v_{xx}D_{y}D_{x}\Lambda))=0.
\end{eqnarray*}
In order to get bilinear B\"acklund transformations of the SUSY BLMP equation, we fix the suitable constraint 
\begin{equation*}
 \mathcal{Y}_{(x,0,0),(0,1)}(v,w)- D_{y}D_{x}\Lambda =D_{y}w_x+v_xD_{y}v- D_{y}D_{x}\Lambda=0.
\end{equation*}
Then one can show, using the above constraint, that
\begin{equation*}
L(\psi)-L(\eta)=-2D_{x}(\mathcal{Y}_{(0,0,t),(0,0)}(v,w)+\mathcal{Y}_{(3x,0,0),(0,0)}(v,w))_y
\end{equation*}
and thus we impose the polynomial constraints
\begin{equation*}
 \mathcal{Y}_{(x,0,0),(0,1)}(v,w)- D_{y}D_{x}\Lambda=0,\, \, \, \mathcal{Y}_{(0,0,t),(0,0)}(v,w)+\mathcal{Y}_{(3x,0,0),(0,0)}(v,w)=0,
\end{equation*} 
which leads, together with the change of variables (\ref{change}), to a set of bilinear transformation
\begin{equation}
 (\mathcal{S}_{y}\mathcal{D}_x- D_{y}D_{x}\Lambda)(\tau\cdot\mu)=0,\quad (\mathcal{D}_t+\mathcal{D}_x^3)(\tau\cdot\mu)=0.\label{bilinear}
\end{equation}
The above transformation is similar to the bilinear form of the SUSY mKdV equation proposed in \cite{LiuHu} identifying the variable $y$ with the variable $x$. It can be viewed as a SUSY generalization of a bilinear Miura transformation.
Furthermore, taking $\tau(x,y,t;\theta)=f(x,y,t)$ and $\mu(x,y,t;\theta)=g(x,y,t)$ in (\ref{bilinear}), one gets the set of bilinear transformations obtained in \cite{Luo} for the classical BLMP equation given by
\begin{equation}
  (\mathcal{D}_{y}\mathcal{D}_x- c^{\prime}(y))(f\cdot g)=0,\quad (\mathcal{D}_t+\mathcal{D}_x^3)(f\cdot g)=0.
\end{equation}

We thus use this transformation to get a set of bilinear B\"acklund transformations of the SUSY BLMP equation. Indeed, we have the following proposition:\\
\textbf{Proposition} Suppose that $(\tau,\mu)$ is a solution of the bilinear equations (\ref{bilinear}), then $(\tau^{\prime},\mu^{\prime})$ satisfying the following bilinear relations
\begin{eqnarray}
 \mathcal{D}_x(\tau\cdot\mu^{\prime})-\alpha\mathcal{D}_x(\mu\cdot\tau^{\prime})=\beta\tau\mu^{\prime}-\alpha\beta\tau^{\prime}\mu,\\
\mathcal{S}_y(\tau\cdot\mu^{\prime})+\alpha\mathcal{S}_y(\mu\cdot\tau^{\prime})=\gamma\tau\mu^{\prime}+\alpha\gamma\tau^{\prime}\mu,\\
(\mathcal{D}_t+\mathcal{D}_x^3-3\beta\mathcal{D}_x^2+3\beta^2\mathcal{D}_x)(\tau\cdot\tau^{\prime})=0,\\
(\mathcal{D}_t+\mathcal{D}_x^3-3\beta\mathcal{D}_x^2+3\beta^2\mathcal{D}_x)(\mu\cdot\mu^{\prime})=0,
\end{eqnarray}
solves the bilinear equations (\ref{bilinear}), where $\alpha$ and $\beta$ are even constants and $\gamma$ is an odd constant.\\
\textit{Proof:} The proof of this proposition is similar to the one presented in \cite{LiuHu}. Indeed, we take
\begin{eqnarray*}
\mathcal{P}_1=2\left(\tau^{\prime}\mu^{\prime}(\mathcal{S}_{y}\mathcal{D}_x- D_{y}D_{x}\Lambda)(\tau\cdot\mu)-\tau\mu(\mathcal{S}_{y}\mathcal{D}_x- D_{y}D_{x}\Lambda)(\tau^{\prime}\cdot\mu^{\prime})\right),\\
\mathcal{P}_2=\tau^{\prime}\mu^{\prime}(\mathcal{D}_t+\mathcal{D}_x^3)(\tau\cdot\mu)+\tau\mu(\mathcal{D}_t+\mathcal{D}_x^3)(\tau^{\prime}\cdot\mu^{\prime}),
\end{eqnarray*}
and show that $\mathcal{P}_1=\mathcal{P}_2=0$ using the above bilinear relations. The proof follows noticing that $\mathcal{P}_1$ is equivalent to
\begin{equation*}
 \mathcal{P}_1=2\left(\tau^{\prime}\mu^{\prime}\mathcal{S}_{y}\mathcal{D}_x(\tau\cdot\mu)-\tau\mu\mathcal{S}_{y}\mathcal{D}_x(\tau^{\prime}\cdot\mu^{\prime})\right).
\end{equation*}

The above B\"acklund transformations associated to the SUSY BLMP equation are related to the ones for the SUSY mKdV equation \cite{LiuHu} and the $\mathcal{N}=2$ SUSY KdV equation \cite{ZLW} when one identifies the variable $y$ with the variable $x$.

\section{Conclusion}
We have started this paper, with an extensive presentation of new solutions of the BLMP equation. Indeed, we have produced using Hirota and Wronskian methods numerous solutions, \textit{e.g} rational, soliton, positon, negaton solutions. Most solutions being introduced for the first time.

We also presented a new supersymmetric Boiti-Leon-Manna-Pempenelli equation. We have studied the SUSY BLMP equation from different view points using the beauty of the SUSY Bell polynomials. The novelty of this paper is the new approach
consisting of finding the superpartners based on solving the linear equation (\ref{Schro})

Futur work should turn around the application of the super Bell polynomials to $\mathcal{N}=2$ extensions of nonlinear PDE's \textit{e.g.} the SUSY KdV equation \cite{LM}.

\subsection*{Acknowledgements}
L. Delisle acknowledges the support of a FQRNT doctoral research scholarship. The authors would like to give special thanks to V\'eronique Hussin for helpful discussions. L. Delisle would also like to thank the department of mathematical sciences of Durham University for its hospitality during the last stage of this paper.

The authors would like to give special thanks to the referees for constructive remarks.

\end{document}